\documentstyle[12pt]{article}

\def\be{\begin{equation}}
\def\ee{\end{equation}}
\def\ba{\begin{array}{c}}
\def\ea{\end{array}}

\def\ben{\[}
\def\een{\]}

\begin{document}

\titlepage
%\vspace*{4cm}

\begin{center}

{\Large \bf Should ${\cal PT}$ symmetric quantum mechanics be
interpreted as nonlinear?}
\end{center}

\vspace{5mm}

 \begin{center}

Miloslav Znojil \vspace{3mm}

\'{U}stav jadern\'e fyziky AV \v{C}R, 250 68 \v{R}e\v{z}, Czech
Republic\\

e-mail: znojil@ujf.cas.cz

\end{center}

\vspace{5mm}

\section*{Abstract}

The Feshbach-type reduction of the Hilbert space to the physically
most relevant ``model" subspace is suggested as a means of a formal
unification of the standard quantum mechanics with its recently
proposed ${\cal PT}$ symmetric modification. The resulting
``effective" Hamiltonians $H_{eff}(E)$ are always Hermitian, and the
two alternative forms of their energy-dependence are interpreted as a
certain dynamical nonlinearity, responsible for the repulsion and/or
attraction of the levels in the Hermitian and/or ${\cal PT}$
symmetric cases, respectively. The spontaneous ${\cal PT}$ symmetry
breaking is then reflected by the loss of the Hermiticity
of~$H_{eff}$ while the pseudo-unitary evolution law persists in the
unreduced Hilbert space.

 \vspace{9mm}

\noindent PACS 03.65.Bz, 03.65.Ca 03.65.Fd

\newpage

\section{\label{I} Introduction}

Bender and Boettcher \cite{BB} tentatively attributed the reality
of spectra in non-Hermitian models to the commutativity of the
Hamiltonians with the product of the complex conjugation ${\cal
T}$ (which mimics the time reversal) and the parity ${\cal P}$,
 \begin{equation}
 H = {\cal PT} H {\cal PT} \equiv  H^\ddagger.
 \label{sebrle}
 \end{equation}
An acceptability of this conjecture is supported by the growing
empirical experience with the similar models \cite{ixnan} and by
the analysis of many examples (\ref{sebrle}) which are partially
\cite{QES} or completely \cite{exact} exactly solvable. In the
physics community, a steady growth of acceptance of the ${\cal
PT}$ symmetric models can be attributed to their phenomenological
relevance in solid state physics \cite{Nelson}, statistical
physics \cite{Zee}, population dynamics \cite{Hatano}, in the
many-body \cite{Calo} and supersymmetric \cite{SUSY} context and,
last but not least, within the general quantum field theory
\cite{BMb}. A reason why the ${\cal PT}$ symmetric models could
eventually prove useful in these applications has been sought in
the reality of their spectrum.

The latter argument is slightly misleading and has been criticized
recently \cite{Japaridze} but the debate proves inspiring,
involves many separate issues and, apparently, may be expected to
continue. In what follows, we intend to join it, emphasizing that
the concept of extended, non-Hermitian quantum mechanics with real
spectra and ${\cal PT}$ symmetric Hamiltonians exhibits multiple
parallels with the standard textbook quantum mechanics.

The material is organized as follows. Firstly, section \ref{III}
reviews several features of the symmetry breaking within the
standard quantum mechanics. It emphasizes that the commutativity
of the Hamiltonian with the parity ${\cal P}$ enables us to split
the Hilbert space into two subspaces. The loss of this
commutativity interrelates these two subspaces but one can still
stay within one of them at a cost of the replacement of the
Hamiltonian $H$ by its Feshbach's \cite{Feshbach} energy-dependent
(so called effective) equivalent form~$H_{eff}(E)$.

In section \ref{3.g} we return to the non-Hermitian models and
stress some of their most important specific properties. In
particular, the reasons for the introduction of an indeterminate
inner product are recollected. We show that in spite of the
non-Hermiticity of $H$, the related pseudo-unitary character of
the time evolution represents a good reason for introduction of a
certain pseudo-norm.

In section \ref{4.11} we formulate the core of our present message
and show that the Feshbach's reduction of the ${\cal PT}$
symmetric operators $H = H^\ddagger$ is Hermitian. As an immediate
consequence, at least a part of the spectrum may be real, and its
full reality may be expected to occur at least in the case of a
certain sufficiently weak non-Hermiticity. Such an idea is also
shown to inspire an immediate generalization of the concept of the
${\cal PT}$ symmetry. An analogy between Hermiticity and ${\cal
PT}$ symmetry is established, in spite of the fact that the
effective Schr\"{o}dinger equation must be understood as slightly
nonlinear. This nonlinearity is ``weak", mediated merely by the
energy-dependence of the reduced Hamiltonians $H_{eff}(E)$.

Section \ref{V} contains the discussion of several related
questions. We pay some attention to the so called spontaneous
${\cal PT}$ symmetry breaking and to the loss of the reality of
the energies, not accompanied by any loss of the pseudo-unitarity
of the time evolution.

Finally, section \ref{VI} summarizes all our results and
emphasizes that the ${\cal PT}$ symmetric quantum mechanics with
real spectra might admit the standard probabilistic physical
interpretation of the wave function on a suitably reduced Hilbert
space.

\section{A short detour to the standard quantum mechanics
\label{III}}

\subsection{A parity-preserving oscillator example}

The quadratic plus quartic one-dimensional Hamiltonian
 \begin{equation}
H(g)=p^2+x^2+g\,x^{4} \label{xN}
 \end{equation}
is extremely popular in perturbation theory where its mathematical
study admits the complex couplings $g$ and, therefore, an explicit
breakdown of Hermiticity. This qualifies this model as a guide
which appears in the standard quantum mechanics~\cite{Simon} as
well as in its ${\cal PT}$ symmetric alternative~\cite{Simsek}.

In the former, purely Hermitian case, all the coupling constants
in eq. (\ref{xN}) are real and  the spectral representation of our
anharmonic oscillator Hamiltonian (which commutes with the parity
${\cal P}$) may be split in the even- and odd-parity eigenstates~$
|\,n^{(\pm)}(g)\rangle $,
 \ben
  H(g) = \sum_{n=0}^{\infty} |\,n^{(+)}(g)\rangle\
\varepsilon^{(+)}_n(g)\, \langle n^{(+)}(g)| + \sum_{m=0}^{\infty}
|\,m^{(-)}(g)\rangle\ \varepsilon^{(-)}_m(g)\, \langle
 m^{(-)}(g)|\ , \ \ \ g \geq 0.
 \een
The parity conservation annihilates some matrix elements between
the $g=0$ basis states denoted by the symbols $|\,s_n\rangle=
|\,n^{(+)}(0)\rangle$ and $|\,t_n\rangle= |\,n^{(-)}(0)\rangle $,
 \begin{equation}
 \langle
s_m|\,H(g)\,|\,t_n\rangle = \langle t_m|\,H(g)\,|\,s_n\rangle = 0,
\ \ \ \ \ m, n = 0, 1, \ldots \ .
 \label{counterpart}
 \end{equation}
This induces the so called super-selection $\langle
s_m|\,\psi^{(-)}\rangle = \langle t_m|\,\psi^{(+)}\rangle =0$ and
splits the variational Schr\"{o}dinger equation in the two
separate infinite-dimensional matrix sub-problems with a definite
parity,
 \ben
  \sum_{k=0}^{\infty}\
 {\cal F}_{mk}\
 \langle s_k|\,\psi^{(+}\rangle
   =E\ \langle s_m|\,\psi^{(+)}\rangle, \ \ \ \ \
  m = 0, 1, \ldots\ , \  \ \ \ \
\
  {\cal F}_{mk} =\langle s_m|\,H(g)\,|\,s_k\rangle,
\label{aboves}
 \een
 \ben
  \sum_{k=0}^{\infty}\
 {\cal G}_{mk}\
 \langle t_k|\,\psi^{(-}\rangle   =E\ \langle
t_m|\,\psi^{(-)}\rangle, \ \ \ \ \
  m = 0, 1, \ldots\ , \  \ \ \ \
\
  {\cal G}_{mk} =\langle t_m|\,H(g)\,|\,t_k\rangle.
\label{abovet}
 \een
This means that the matrix form of $H(g)$ is the direct sum of the
two {different} matrices ${\cal F}$ and ${\cal G}$, precisely in
the spirit of the Schur's lemma~\cite{Blank}.

\subsection{Parity-breaking terms and the effective Hamiltonians}

For a more general anharmonic oscillator
 \ben
 H(f,g)=p^2+x^2+f\,x^{3}+g\,x^{2N}
 \een
the parity ${\cal P}$ ceases to be a useful symmetry due to the
presence of the spatially asymmetric cubic term whose
non-vanishing elements form a matrix $\Omega_{mj}=\langle t_m
|\,f\,x^{3}\, |\,s_j\rangle$. Each wave function in $L_2(I\!\!R)$
must be expanded in the full basis,
 \begin{equation}
 |\psi\rangle =
  \sum_{n=0}^{\infty}
 |s_n\rangle\
 h^{(+)}_n
 + \sum_{m=0}^{\infty}
 |t_m\rangle\
 h^{(-)}_m\ .
 \label{eded}
 \end{equation}
Schr\"{o}dinger equation acquires the partitioned matrix form
 \ben
 \left (
 \begin{array}{cc}
{\cal F} -E\,I & \Omega^T\\
 \Omega&{\cal G} -E\,I
 \ea
 \right )\,
 \left (
 \ba
 \vec{h}^{(+)}\\
 \vec{h}^{(-)}
 \ea
 \right ) = 0
 \label{StEre}
 \een
where we may eliminate the Feshbach's \cite{Feshbach}
``out-of-the-model-space" components
 \ben
 \vec{h}^{(-)} = -\frac{1}{{\cal G}
 -E\,I}\,
 \Omega\,\vec{h}^{(+)}
 \label{elimi}
 \een
and get the reduced Schr\"{o}dinger equation containing the
effective Hamiltonian which is (presumably, not too manifestly)
energy dependent,
 \begin{equation}
  H_{eff}(E)\,\vec{h}^{(+)} =  E\,\vec{h}^{(+)},\ \ \ \ \ \ \ \
 H_{eff}(E)=\left (
 {\cal F}-
\Omega^T\,
 \frac{1}{{\cal G}-E\,I}\,
\Omega
 \right )\,.
 \label{StEfin}
 \end{equation}
The energy-dependence of $H_{eff}(E)$ causes rarely a problem. In
numerical context one fixes a trial energy $\varrho$ in
$H_{eff}(\varrho)$ and solves the linearized Schr\"{o}dinger
equation giving a one-parametric family of auxiliary spectra
$\{\hat{E}_n(\varrho)\}$. A return to the exact and nonlinear
eigenvalue problem (\ref{StEfin}) is then mediated by the
selfconsistent determination of the best parameter,
 \begin{equation}
\varrho =\hat{E}_n(\varrho).
 \label{selfco}
 \end{equation}
Sufficient precision is mostly achieved via the linear
approximation
 \begin{equation}
  H_{eff}(\varrho)\,\vec{h}^{(+)} =  E\,\vec{h}^{(+)},\ \ \ \ \ \ \ \
 \label{StEfinta}
 \end{equation}
with a single value of $\varrho$ adapted to the practical
evaluation of a set of several neighboring energy levels~$E$.

\section{\label{3.g} ${\cal PT}$ symmetric formalism}

% \subsection{A concise outline of history \label{II}}

An interest in the commutativity (\ref{sebrle}) of $H$ with ${\cal
PT}$ (let us repeat that ${\cal P}$ means parity and ${\cal T}$
denotes time reversal) grew from several sources.  The oldest root
of its appeal is the Rayleigh-Schr\"{o}dinger perturbation theory.
Within its framework, Caliceti et al \cite{Caliceti} have
discovered that a low-lying part of the spectrum in the cubic
anharmonic potential $V= x^2 + g\,x^3$ for some {\em purely
imaginary} couplings $g$ is {\em real}.  This establishes an
analogy between the Hermitian and some non-Hermitian oscillators,
extending the family of the eligible phenomenological potentials.

A non-perturbative direction of analysis has been initiated by
Buslaev and Grecchi \cite{BG} who were motivated by the physical
relevance of non-Hermitian models in field theory. They employed
parallels between Hermiticity and ${\cal PT}$ symmetry during
their solution of an old puzzle of spectral equivalence between
apparently non-equivalent quartic interaction models.  Bender and
Milton \cite{QED} underlined in similar context that an ambiguity
in boundary conditions exists and is essential for the
clarification and consequent explanation of the famous Dyson's
paradox in QED. These studies opened new mathematical as well as
interpretation problems. Some of them will be discussed here.

\subsection{\label{2.1} Modified inner product}

In ${\cal PT}$ symmetric quantum mechanics the Hamiltonians are
non-Hermitian and one often discovers (or, at worst, assumes)
that their spectrum is real, discrete and non-degenerate.  Even
under this assumption, their left eigenvectors (let us denote
them by the symbol $\langle \langle \psi|$) need not necessarily
coincide with the Hermitian conjugates $\langle \psi|$ of their
right eigenvector partners. The Hermitian conjugation must be
replaced by its modification,
 \ben
\langle \psi | \to \langle \langle \psi|= \langle \psi |\, {\cal
P} \ .
 \label{recep}
 \een
Originally, such a replacement has been made and used in the
non-degenerate perturbation theory \cite{mytri} where the
Rayleigh-Schr\"{o}dinger formalism leads to the recursive
definition of the products $E^{(k)} \cdot \langle \psi^{(0)}|\,
{\cal P} |\, \psi^{(0)} \rangle $. They contain the energy
correction $E^{(k)} $ multiplied by the unperturbed pseudo-norm
which vanishes precisely at the boundary of the applicability of
the non-degenerate perturbation formalism. At this boundary a {\em
real} Bender-Wu singularity is crossed~\cite{Alvarez} so that
$\langle \psi^{(0)}|\, {\cal P} |\, \psi^{(0)} \rangle \to 0$ and
a pair of the energy levels merges~\cite{ptho,Morse,Coulomb}.

At the two different real energies $E_1 \neq E_2$ the comparison
of the left and right ${\cal PT}$ symmetric equations
$H|\psi_1\rangle =E_1|\psi_1\rangle$ and $\langle \langle
\psi_2|\,H = \langle \langle \psi_2|\,E_2$ leads to the
orthogonality $\langle \langle \psi_2|\, \psi_1\rangle=0$ so that
the inner product with metric ${\cal P}$ and with the so called
quasi-parity $Q_n= \pm 1$ \cite{ptho,Morse} is the natural option.
Formally, the disappearance of the self-overlap $\langle \psi |
{\cal P} | \psi \rangle =0$ does not imply that the vector $ |
\psi \rangle$ itself must vanish so that the requirement
 \begin{equation}
\langle \psi_n | {\cal P}| \psi_m \rangle = Q_n \delta_{mn}, \ \ \
\ \ \ m , n = 0, 1, \ldots .
 \label{rumor}
 \end{equation}
merely ``pseudo-normalizes" the solutions (cf. also ref.
\cite{Quesne}). A further development of the theory requires the
notion of the completeness of the bound states,
 \ben
\sum_{n=0}^\infty | \psi_n \rangle \, Q_n \langle \psi_n | {\cal
P} =  I
 \een
as well as an innovated spectral representation of a given
non-Hermitian ${\cal PT}$ symmetric Hamiltonian with real
spectrum,
 \ben
H= \sum_{n=0}^\infty | \psi_n \rangle \, E_n\,Q_n \langle \psi_n |
{\cal P}.
 \een
It admits various pseudo-Hermitian alternatives and
generalizations~\cite{Mostafazadeh}.

\subsection{\label{3.2} The pseudo-unitarity of the evolution in time}

Evolution of bound states in quantum mechanics is mediated
(generated) by their Hamiltonian,  $
 |\psi[t]\rangle = \exp (-i\,H\,t)\,
 |\psi[0]\rangle$.
In the models with Hermitian $H =H^\dagger$ the availability of
solutions of the time-independent Schr\"{o}dinger equation
% $ H\,|\psi_n\rangle =
%  E_n\,|\psi_n\rangle$
simplifies this rule since all the eigenvalues $E_n$ remain real
and the time-dependence of the separate eigenstates becomes
elementary,
 \ben
 |\psi_n[t]\rangle = e^{-i\,E_n\,t}\,
 |\psi_n[0]\rangle\ .
 \label{dva}
 \een
Although a fully consistent and complete physical interpretation
of the general pseudo-Hermitian Hamiltonians is not at our
disposal yet, many of their formal features are not entirely new,
mimicking the models with indefinite metric in relativistic
physics etc. Another significant source of insight are particular
examples. In many of them, whenever the real energies $E_n$ are
attributed to a non-Hermitian, ${\cal PT}$ symmetric  Hamiltonian
with the property (\ref{sebrle}), we may infer that
 \ben
 |\psi[t]\rangle = e^{-iHt}\,
 |\psi[0]\rangle
=
\sum_{n=0}^\infty | \psi_n \rangle \,e^{-i\, E_n\,Q_n\,t} \,
\langle \psi_n | {\cal P}
 |\psi[0]\rangle .
 \een
This formula means that the conservation law concerns the
innovated scalar product,
 \ben
\langle \psi[t]| {\cal P}| \psi[t] \rangle= \langle \psi[0]| {\cal
P}| \psi[0] \rangle
 \een
so that the time evolution of the system is pseudo-unitary.

\section{\label{4.11} An explanation of the reality of spectra}

Any eigenstate of $H = H^\dagger ={\cal P}H{\cal P}$ (e.g., of
$H(g)$ in paragraph 2.1) satisfies the same Schr\"{o}dinger
equation even when it is pre-multiplied by the parity ${\cal P}$.
Both the old and new eigenstates belong to the same real
eigenvalue $E$ which cannot be degenerate due to the
Sturm-Liouville oscillation theorems. One of the superpositions
$|\psi\rangle \pm {\cal P}|\psi\rangle$ must vanish while the
other one acquires a definite parity.  This is the essence of the
mathematical proof of the above-mentioned Schur's lemma. The wave
functions are even or odd and the ${\cal P}$ symmetry of wave
functions cannot be spontaneously broken, ${\cal P} |n^{(\pm)}(g)
\rangle= \pm |n^{(\pm)}(g) \rangle$.

The rigidity of the latter rule is lost during the transition to
the ${\cal PT}$ symmetric models where any quantity
$\exp(i\varphi)$ is an admissible eigenvalue of the operator
${\cal PT}$  since its component ${\cal T}$ is defined as
anti-linear, ${\cal T}i=-i$. In more detail, every rule ${\cal
PT}|\psi\rangle = \exp(i\varphi)\,|\psi\rangle$ implies that we
have
 \ben
 {\cal PT}{\cal PT}|\psi\rangle =
\exp(-i\varphi)\,{\cal PT}|\psi\rangle= |\psi\rangle
 \een
as required. The Schur's lemma ceases to be applicable. In the
basis with the properties ${\cal PT} |S\rangle =
|S\rangle$ and ${\cal PT} |L\rangle = -|L\rangle$, the general
expansion formula
 \ben
 H= \sum_{m,n=0}^{\infty}  \left (
 \ba
 \\
 \ea
 |S_m\rangle {\cal F}_{m,n} \langle S_n | \,+\,
 |L_m\rangle {\cal G}_{m,n} \langle L_n | \,+\,
 i\,|S_m\rangle {\cal  C}_{m,n} \langle L_n | \,+\,
 i\,|L_m\rangle {\cal D}_{m,n} \langle S_n |\
 \right )
 \een
contains four separate complex matrices of coefficients.  Once it
is subdued to the requirement $H= {\cal PT} H{\cal PT}$, we get
the necessary and sufficient condition demanding that all the
above matrix elements of $H=H^\ddagger$ must be real,
 \begin{equation}
 {\cal F}_{m,n} = {\cal F}_{m,n}^*, \ \ \ \ \  {\cal G}_{m,n}
= {\cal G}_{m,n}^*, \ \ \ \ \   {\cal C}_{m,n} = {\cal C}_{m,n}^*,
\ \ \ \ \ {\cal D}_{m,n} = {\cal D}_{m,n}^*.
 \label{constrainte}
 \end{equation}
As long as the similar trick has led to the superselection rules
for the spatially symmetric Hamiltonians, we may conclude that the
${\cal PT}$ symmetric analogue of the direct-sum decompositions
and superselection rules (\ref{counterpart}) is just the much
weaker constraint~(\ref{constrainte}).

\subsection{\label{IV} Re-emergence of Hermiticity via effective
Hamiltonians}

Whenever we have a state with the ${\cal PT}-$parity equal to
$\exp(i\varphi)$, we may try to shift the phase and introduce the
new state $|\chi\rangle= \exp(i\beta) |\varphi\rangle$. The action
$ {\cal PT}\,|\varphi\rangle= e^{i\varphi}|\varphi\rangle$ is
modified,
 \ben
 {\cal PT}\,|\chi \rangle=  {\cal PT}\,e^{i\beta}|\varphi\rangle
= e^{-i\beta} {\cal PT}\,|\varphi\rangle = e^{i\,(\varphi-2\beta)}
\,|\chi\rangle
 \een
and the ${\cal PT}$ parity has changed by $-2\beta$. Via the
renormalization $\beta$ we may achieve that the new ${\cal PT}$
phase is zero. Such a normalization convention means that
  \ben
 |\psi\rangle =
  \sum_{n=0}^{\infty}
 |s_n\rangle\
 p^{(+)}_n
 +i\, \sum_{m=0}^{\infty}
 |t_m\rangle\
 p^{(-)}_m\
 \label{peded}
 \een
where all the coefficients are real. This revitalizes the analogy
with formula (\ref{eded}). Our next illustration,
 \ben
H(if,g)=p^2+x^2+if\,x^{3}+g\,x^{4}
 \een
may use the {\em same} matrix elements $\Omega_{mj}=\langle t_m
|\,f\,x^{3}\, |\,s_j\rangle$ as above and becomes tractable by the
mere replacements $\Omega \to i\,\Omega$, $h^{(+)}_n \to
p^{(+)}_n$  and $h^{(-)}_n \to i\,p^{(-)}_n$. This gives very
similar, real Schr\"{o}dinger matrix equation
 \ben
 \left (
 \begin{array}{cc}
{\cal F} -E\,I & - \Omega^T\\
 \Omega&{\cal G} -E\,I
 \ea
 \right )\,
 \left (
 \ba
 \vec{p}^{(+)}\\
 \vec{p}^{(-)}
 \ea
 \right ) = 0
 \label{StErep}
 \een
and the very similar partial solution
 \ben
 \vec{p}^{(-)} = +\frac{1}{{\cal G}
 -E\,I}\,
 \Omega\,\vec{p}^{(+)}.
 \label{elimip}
 \een
We have to emphasize that the final, effective Schr\"{o}dinger
equation is Hermitian,
 \begin{equation}
  H_{eff}(\varrho)\,\vec{p}^{(+)} =  E(\varrho)\,\vec{p}^{(+)},\ \ \ \ \ \ \ \
 H_{eff}(\varrho)=\left (
 {\cal F}+
\Omega^T\,
 \frac{1}{{\cal G}-\varrho\,I}\,
\Omega
 \right )\,
 \label{StEfinp}
 \end{equation}
In comparison with the recipe of paragraph 3.2 the only difference
is in the sign of the correction term.  This makes the connection
between the Hermiticity and ${\cal PT}$ symmetry particularly
tight.  Both the Schr\"{o}dinger equations (\ref{StEfin}) and
(\ref{StEfinp}) are Hermitian and give the (different) real
spectra $E(\varrho)$ at any $\varrho$. Both these reduced
Schr\"{o}dinger equations prove insensitive to the change of the
sign of the coupling matrix $\Omega$ but a return to the original
Schr\"{o}dinger equations reveals that the replacement $ \Omega
\to - \Omega$ is not an equivalence transformation as it changes
the wave functions.

\subsection{The generalized metric operators ${\cal  P}$}

One has to impose the selfconsistency condition (\ref{selfco}) but
it is clear that this cannot give any complex roots $E(E)$ in the
Hermitian case. In contrast, they may freely emerge in the
non-Hermitian setting so that the ${\cal PT}$ symmetry is less
robust than Hermiticity.

All the other parallels between the Hermitian an ${\cal PT}$
symmetric models are more straightforward. Once we work just with
the effective Hamiltonian (which is always Hermitian), many
phenomenologically oriented conclusions concerning the Hermiticity
or pseudo-Hermiticity of the full, original Hamiltonian will only
depend on the subtle details of an overall energy or rather
$\varrho-$dependence of our model-space Hamiltonian
$H_{eff}(\varrho)$. In this sense, the ${\cal PT}$ symmetry may be
considered to be just a very special case of the
pseudo-Hermiticity.

Let us now return to our original problem once more. Why do the
${\cal PT}$ symmetric Hamiltonians have real energies? The above
explanation relies on the Hermiticity of $H_{eff}(\varrho)$,
guaranteeing that all the auxiliary $E_n(\varrho)$ are real. The
discussion is reduced to the selfconsistency (\ref{selfco}) and to
the reality/complexity of its roots. In this sense the whole
parallel between the Hermitian and non-Hermitian coupling of the
individual sub-Hamiltonians ${ F}$ and ${ G}$ is based on the mere
matrix structure of the Schr\"{o}dinger equation. Its partitioned
form
 \begin{equation}
\left (
\begin{array}{cc}
F-E\,I&\alpha\,A\\ A^\dagger& G-E\,I
 \ea
\right ) \cdot \left ( \ba \vec{u}\\ \vec{w}
 \ea
\right ) = 0\ \label{SEdrd}
 \end{equation}
represents simply the Hermitian case at $\alpha=1$ and the ${\cal
PT}$ symmetric case at $\alpha=-1$. Thus, the operator ${\cal P}$
need not be parity. As long as our previous analysis did not
depend on this interpretation, the real energies may be expected
to emerge, following the same idea of the effective Hermitization,
from the other matrix structures of $H$.

We may admit, for example, that the block $A$ is not a real matrix
at all. One can imagine that the {\em complex} (and Hermitian or
even merely ${\cal PT}$ symmetric) sub-Hamiltonians $F$ and $G$
would also lead to the real spectra, at least in the limit of the
sufficiently small {\em complex} coupling matrices~$A$.

Another type of generalization was present in the original
Feshbach's proposal \cite{Feshbach} where the upper partition $F$
is the most relevant part of the Hilbert space (called ``model"
space) spanned just by a few most important elements of the basis.
The other partition is usually expected to contribute to the
observable quantities as a correction. Thus, one might work with
the two partitions of different size, ${\rm dim}\,F \neq {\rm
dim}\,G$.

Last but not least, one could consider a triple or multiple
partitioning which would generalize eq. (\ref{SEdrd}). An explicit
construction of this type may be found, e.g., in our recent
remark~\cite{Krakov}.

\section{Discussion \label{V}}

\subsection{What happens after a spontaneous breakdown
of ${\cal PT}$ symmetry}

A puzzle emerges when the non-Hermiticity grows and certain
doublets of real energies merge and form, subsequently, complex
conjugate pairs. Explicit examples of such a possibility range
from the ${\cal PT}$ symmetric square well on a finite interval
\cite{sqw} up to many quasi-exactly solvable models \cite{Ahmed}
and virtually all the shape invariant potentials on the whole real
line \cite{Geza}.

Let us recollect the harmonic oscillator example $H =
p^2+r^2+G/r^2$ of ref. \cite{ptho} and the two possible forms of
its energy spectrum.  At $G> -{1}/{4}$ one encounters the purely
real and discrete levels $E_N=4n+2-2\,Q\,\gamma$ with $\gamma =
\sqrt{-G-{1}/{4}}>0$ and $n = 0, 1, \ldots$. These levels
(distinguished by their quasi-parity $Q = \pm 1$) are to be
compared with the complex conjugate pairs $E_N=4n+2-2\,i\,
Q\,\delta$ which replace the above set at the strongly negative
coupling $G < -{1}/{4}$ in $\delta = \sqrt{-G-{1}/{4}}>0$. We see
that once we remove the constraint $G>-1/4$, the ${\cal PT}$
symmetry of the wave functions breaks down {\em for all the levels
at once}, at $G = -1/4$~\cite{Geza}.

In such a context let us now assume that the solution of a given
non-Hermitian Schr\"{o}dinger equation gives at least two energies
which are mutual complex conjugates,
 \begin{equation}
H |\psi_+\rangle = E\,|\psi_+ \rangle, \ \ \ \ \ \ \ \ \ \ \ H
|\psi_{-}\rangle = E^*\,|\psi_{-} \rangle.
 \label{conjs}
 \end{equation}
We may re-write these two Schr\"{o}dinger equations in their
respective Hermitian conjugate form with $H^\dagger={\cal P} \,H
\,{\cal P}$ acting to the left,
 \ben
\langle \psi_+ | \,{\cal P} \,H  = E^*\, \langle \psi_+ |\,{\cal
P}, \ \ \ \ \ \ \ \ \ \ \ \langle \psi_{-} | \,{\cal P} \,H  = E\,
\langle \psi_{-} |\,{\cal P}.
 \een
Out of all the possible resulting overlaps, let us now recollect
the following four,
 \ben
\langle \psi_{+} |\,{\cal P}\, H |\psi_{+}\rangle = E^*\,\langle
\psi_{+} |\,{\cal P}\, |\psi_{+} \rangle, \ \ \ \ \ \ \ \ \ \ \
\langle \psi_{+} | \,{\cal P} \,H\,|\psi_{+} \rangle  = E\,
\langle \psi_{+} |\,{\cal P}\,|\psi_{+} \rangle ,
 \een
 \ben
\langle \psi_{-} |\,{\cal P}\, H |\psi_{-}\rangle = E^*\,\langle
\psi_{-} |\,{\cal P}\, |\psi_{-} \rangle, \ \ \ \ \ \ \ \ \ \ \
\langle \psi_{-} | \,{\cal P} \,H\,|\psi_{-} \rangle  = E\,
\langle \psi_{-} |\,{\cal P}\,|\psi_{-} \rangle.
 \een
Their comparison suggests that for $E\neq E^*$ the self-overlaps
must vanish,
 \ben
\langle \psi_+ |\,{\cal P}\, |\psi_+\rangle = 0, \ \ \ \ \ \ \ \ \
\ \ \langle \psi_{-} |\,{\cal P}\, |\psi_{-}\rangle = 0.
 \een
We must extend the above rule (\ref{rumor}) and complement it by
an off-diagonal pseudo-normalization
 \begin{equation}
\langle \psi_+ |\,{\cal P}\, |\psi_{-}\rangle = \left [ \ \langle
\psi_{-} |\,{\cal P}\, |\psi_{-}\rangle \ \right ]^*= c
 \label{offnorm}
 \end{equation}
with any suitable $c \in l\!\!\!C$.

%  2.3.  The conservation of pseudo-norm (3.2,b and 4.3)

\subsection{The pseudo-norm and its conservation }

For the sake of simplicity let us assume that the ${\cal PT}$
symmetry is broken just at the two lowest states (cf. the examples
\cite{sqw,Simsek}). Besides the above-mentioned modification of
the orthogonality relations,  one has to change the first two
terms in the decomposition of unit,
 \ben
I = | \psi_+ \rangle \, \frac{1}{c^*}\, \langle \psi_{-} | {\cal
P}\  + \ | \psi_{-} \rangle \, \frac{1}{c}\, \langle \psi_+ |
{\cal P}\ + \ \sum_{n=2}^\infty | \psi_n \rangle \, Q_n \langle
\psi_n | {\cal P}  .
 \een
This is a new form of the completeness relations. The parallel
spectral decomposition of the Hamiltonian in question contains the
similar two new terms,
 \ben
H = | \psi_+ \rangle \, \frac{E}{c^*}\, \langle \psi_{-} | {\cal
P}\  + \ | \psi_{-} \rangle \, \frac{E^*}{c}\, \langle \psi_+ |
{\cal P}\ +\ \sum_{n=2}^\infty | \psi_n \rangle \,E_n Q_n \langle
\psi_n | {\cal P}  .
 \een
Finally, the pseudo-unitary time dependence of wave functions
acquires the following new compact form,
 \ben
 |\psi[t]\rangle = e^{-iHt}\,
 |\psi[0]\rangle
=
 \left (
 | \psi_+ \rangle \, \frac{1}{c^*}\, e^{-i\,E\,t}\, \langle
\psi_{-} | {\cal P}\ \right ) + \left ( \ | \psi_{-} \rangle \,
\frac{1}{c}\, e^{-i\,E^*\,t}\, \langle \psi_+ | {\cal P}\right )\
+ \een
 \ben
 \
+\ \sum_{n=2}^\infty | \psi_n \rangle \,e^{-i\, E_n\,Q_n\,t} \,
\langle \psi_n | {\cal P}
 |\psi[0]\rangle
 .
 \een
The value of the scalar product is conserved in time,
 \ben
\langle \psi[t]| {\cal P}| \psi[t] \rangle= \langle \psi[0]| {\cal
P}| \psi[0] \rangle.
 \een
A weakened form of the Stone's theorem could be re-established for
the pseudo-unitary evolution allowing non-Hermitian Hamiltonians
$H = H^\ddagger$.  We see that this may be done not only in the
${\cal PT}$ symmetric systems characterized by the real spectra
but also in the domain of couplings where this symmetry is
spontaneously broken.  A parallel to the unbroken case is
established.  As long as the vanishing self-overlaps $\langle
\langle \psi | \psi \rangle=0$ cease to carry any information
about the phase and scaling of $|\psi\rangle$, the complexified
pseudo-norm may be re-introduced via the off-diagonal rule
(\ref{offnorm}) if needed. In the light of eqs. (\ref{conjs})
which indicate that we may choose $|\psi_-\rangle = {\cal PT}
|\psi_+\rangle$, we may drop the unit operators ${\cal P}^2=1$
from all the overlaps and conclude that our definition of the
inner product should be rewritten in the form
 \begin{equation}
\langle \langle \psi | \psi' \rangle=
 \langle \psi | \vec{\cal T} |\psi' \rangle
 \end{equation}
where the superscripted arrow indicates that the antilinear
operator ${\cal T}$ should be understood as acting,
conventionally, to the right. This makes this  more universal
definition a bit clumsy. Fortunately, whenever the ${\cal PT}$
symmetry is not broken, this new prescription is equivalent to the
old one and can replace it in the orthogonality relations
(\ref{rumor}) etc.

\section{\label{VI} Summary}

In the current literature we are witnessing an increase of
interest in the non-Hermitian Hamiltonians exhibiting ${\cal
PT}$ symmetry and combining promising features (e.g., a
``non-robust" existence of real spectra) with several unanswered
questions.  We motivated our present considerations, mainly, by
the apparent lack of any clear probabilistic interpretation of
wave functions.

Mathematically, it is reflected by the non-unitarity of the time
evolution and by the concept of quasi-parity $Q=\pm 1$ introduced
via a few examples and specified as a certain ``analytic
continuation" of the ordinary quantum number of parity.  On the
spiked harmonic oscillator we illustrated its role of a physical
criterion which distinguishes between the quasi-odd and quasi-even
solutions. In a parallel to the Hermitian picture we eliminated
the latter states from the ``relevant" Hilbert space using the
standard Feshbach projection method.

A formal support for the latter conjectured transition $H \to
H_{eff}$ may be seen in a necessity of suppression of the
indeterminate character of the pseudo-norm within physical space.
This has been amply rewarded. A deep connection between the
Hermitian and ${\cal PT}$ symmetric $H$ has been found in the
shared Hermiticity of their projected forms $H_{eff}$.
The $H_{eff}$ of the respective Hermitian and ${\cal
PT}$ symmetric origin differs just by the sign $\alpha = \pm 1$ of
the correction term.

We hope that we have answered our original question: The
non-Hermitian ${\cal PT}$ symmetric quantum mechanics seems to
find, in its specific and Hermitian projected form, a fairly
natural interpretation.
We have reached a new level of understanding of what happens in
the non-Hermitian systems. There seems to exist a certain
natural boundary of the domain ${\cal D}$ of parameters in $H$.
In its interior the energies stay real.  In the other words, the
``non-obligatory" ${\cal PT}$ symmetry of the wave functions
themselves becomes (people usually say spontaneously) broken on
the boundary of ${\cal D}$. The algebraic manifestation of the
crossing of this boundary (the pseudo-norm vanishes) is
reflected by the disappearing roots in the selfconsistency or
graphical rule~(\ref{selfco}).

\section*{Acknowledgement}

Supported by GA AS (Czech Republic), grant Nr. A 104 8004.

% \newpage

% \def\BibTeX{{\rm B\kern-.05em{\sc i\kern-.025em b}\kern-.08em
  %  T\kern-.1667em\lower.7ex\hbox{E}\kern-.125emX}}

%\label{lastpage}


\begin{thebibliography}{99}
\small


\bibitem{BB}
Bender C M and Boettcher S, Phys. Rev. Lett. { 24} (1998), 5243.

\bibitem{ixnan}
Hatano N and Nelson D R, Phys. Rev. Lett. 77 (1996), 570;

Delabaere E and Pham F, Phys. Lett. A 250 (1998), 25;

Bender C M, Boettcher S and Meisinger P N, J. Math. Phys. 40
(1999), 2201;

L\'evai G, Cannata F and Ventura A,
                 J. Phys. A: Math. Gen.  34 (2001), 839;

Dorey P, Dunning C and Tateo R, J. Phys. A: Math. Gen.  34 (2001),
5679.

\bibitem{QES}
Bender C M and Boettcher S, J. Phys.{ A: Math. Gen. 31} (1998),
L273;

Znojil M, J. Phys.{ A: Math. Gen. 32} (1999), 4563;

Bagchi B, Cannata F and Quesne C, Phys. Lett. A 269 (2000), 79;

Znojil M, J. Phys.{ A: Math. Gen. 33} (2000), 6825;

Bagchi B and  Quesne C, Phys. Lett. A 273 (2000),  285;

Bagchi B, Mallik S, Quesne C and Roychoudhury R, Phys. Lett. A 289
(2001), 34.

\bibitem{exact}
Cannata F, Junker G and Trost J, Phys. Lett. { A 246} (1998), 219;

Znojil M, J. Phys.{ A: Math. Gen. 33} (2000), 4561;

L\'evai G and Znojil M, J. Phys.{ A: Math. Gen. 33} (2000), 7165.

\bibitem{Nelson}
Hatano N and Nelson D R, Phys. Rev. B 56 (1997), 8651.

\bibitem{Zee}
Feinberg J and Zee A, Phys. Rev. E 59 (1999), 6433.

\bibitem{Hatano}
Nelson D R and Shnerb N M 1998 Phys. Rev. E 58 (1999), 1383;

Shnerb N M, Phys. Rev. E 63 (2001), 011906.

\bibitem{Calo}
Znojil M and Tater M, J. Phys. A: Math. Gen. 34 (2001), 1793;

Basu-Mallick B and Mandal B P, Phys. Lett. A 284 (2001), 231.

\bibitem{SUSY}
Andrianov A A, Cannata F, Dedonder J P and Ioffe M V, Int. J. Mod.
Phys. A 14 (1999), 2675;

Znojil M, Cannata F, Bagchi B and Roychoudhury R, Phys. Lett. B
483 (2000), 284;

Klishevich S M and Plyushchay M S, Nucl. Phys. B 606 (2001), 583;

Cannata F, Ioffe M, Roychoudhury R and Roy P, Phys. Lett. A 281
(2001), 305;

Dorey P, Dunning C and Tateo R, J. Phys. A: Math. Gen.  34 (2001),
L391.

\bibitem{BMb}
Grignani G,  Plyushchay M and  Sodano P, Nucl.Phys. B464 (1996),
189;

Bender C M and Milton K A, Phys. Rev. D 55 (1997), R3255;

Nirov K and Plyushchay M,  Nucl.Phys. B512 (1998), 295;

Bender C M and Milton K A, Phys. Rev. D 57 (1998), 3595;

Mostafazadeh A, J. Math. Phys. 39 (1998), 4499.

\bibitem{Japaridze}
Japaridze G S, ``Space of state vectors in PT symmetrized quantum
mechanics", arXiv: quant-ph/0104077;

Kretschmer R and Szymanowski L, ``The interpretation of
quantum-mechanical models with non-Hermitian Hamiltonians and real
spectra", arXiv: quant-ph/0105054.

\bibitem{Feshbach}
Feshbach H, Ann. Phys. (N.Y.) 5 (1958), 357.

\bibitem{Simon}
Simon B, Ann. Phys. (NY) 58 (1970), 76;

Sk\'{a}la L, \v{C}\'{\i}\v{z}ek J and Zamastil J, J. Phys. A:
Math. Gen 32 (1999), 5715.

\bibitem{Simsek}
Znojil M, J. Phys.{ A: Math. Gen. 32} (1999), 7419;

Bender C M, Berry M, Meisinger P N, Savage V M and Simsek M, J.
Phys. A: Math. Gen. 34 (2001), L31.

\bibitem{Blank}
Blank J, Exner P and Havl\'{\i}\v{c}ek M, Hilbert Space Operators
in Quantum Physics, AIP, New York, 1994, p. 244.

\bibitem{Caliceti}
Calicetti E, Graffi S and  Maioli M, Commun. Math. Phys. 75
(1980), 51.

\bibitem{BG}
Buslaev V and Grecchi V, J. Phys.{ A: Math. Gen. 26} (1993), 5541.

\bibitem{QED}
Bender C M and Milton K A, J. Phys. A: Math. Gen. 32 (1999), L87.

\bibitem{mytri}
Fernandez F, Guardiola R, Ros J and Znojil M, J. Phys.{ A: Math.
Gen. 31} (1998), 10105.

\bibitem{Alvarez}
Bender C M and Wu T T, Phys. Rev. 184 (1969), 1231;

Alvarez G, J. Phys. A: Math. Gen. 27 (1995), 4589.

\bibitem{ptho}
Znojil M, Phys. Lett. A. 259 (1999), 220.

\bibitem{Morse}
Znojil M, Phys. Lett. { A 264} (1999), 108.

\bibitem{Coulomb}
Znojil M and L\'{e}vai G, Phys. Lett. { A 271} (2000), 327.

\bibitem{Quesne}
Znojil M, ``What is PT symmetry?", arXiv, quant-ph/0103054, 2001;

Bagchi B, Quesne C and Znojil M, Mod. Phys. Lett. A 16 (2001),
2047.

\bibitem{Mostafazadeh}
Znojil M 2001 ``Conservation of pseudonorm in PT symmetric quantum
mechanics", arXiv, math-ph/0104012, 2001;

Mostafazadeh A, J. Math. Phys. 43 (2002), 205.

\bibitem{Krakov}
Znojil M 2001 ``A generalization of the concept of PT symmetry",
[contrb. QTS2 int. conf. (Krakow, Poland, July 18 - 21), to appear
in proceedings], arXiv, math-ph/0106021,  2001.

\bibitem{sqw}
Znojil M, Phys. Lett. A 285 (2001), 7.

\bibitem{Ahmed}
Ahmed Z,  Phys. Lett. A 282 (2001), 343 and 286 (2001), 231;

Handy C R, Khan D, Wang X-Q and Tomczak C J, J. Phys. A: Math.
Gen. 34 (2001), 5593.

\bibitem{Geza}
L\'{e}vai G and Znojil M, Mod. Phys. Lett. A 16 (2001), 1973.

\end{thebibliography}
\end{document}